\documentclass{ws-rv9x6}
\usepackage{subfigure}   
\usepackage{ws-rv-thm}   
\usepackage{ws-rv-van}   
\makeindex

\begin{document}

\chapter[Badis Ydri]{The QM/NCG Correspondence}

\author[Badis Ydri]{Badis Ydri}

\address{Department of Physics, Badji-Mokhtar Annaba University,\\
 Annaba, Algeria.\\
 ydribadis@gmail.com\\
{\it In Honor of Bal on the Occasion of His 85th Birthday}}

\begin{abstract}
The correspondence between quantum mechanics and noncommutative geometry is illustrated in the context of the noncommutative ${\rm AdS}^2_{\theta}/{\rm CFT_1}$  duality where ${\rm CFT}_1$ is identified as conformal quantum mechanics. This model is conjectured here to describe the gauge/gravity correspondence in one dimension. 
\end{abstract}


\body


\section{A ${\rm QM}/{\rm NCG}$ Correspondence}

In this article we will focus on the case of two dimensions with Euclidean signature where the positive curvature space is given by a sphere $\mathbb{S}^2$ with isometry group $SO(3)$ and the negative curvature space is given by a pseudo-sphere $\mathbb{H}^2$  with isometry group $SO(1,2)$ (which is Euclidean ${\rm AdS}^2$). The space $\mathbb{S}^2\times {\rm AdS}^2$ is in fact the near-horizon geometry of extremal black holes in general relativity and string theory. The information loss problem in  four dimensions on  $\mathbb{S}^2\times {\rm AdS}^2$ is then reduced to the information loss problem in two dimensions on  ${\rm AdS}^2$.

The quantization of these two spaces yields the fuzzy sphere $\mathbb{S}^2_N$  \cite{Hoppe-Madore:1991bw} and the noncommutative pseudo-sphere ${\rm AdS}^2_{\theta}$ \cite{Ho:2000fy,Jurman:2013ota,Pinzul:2017wch} respectively which enjoy the same isometry groups $SO(3)$ and $SO(1,2)$ as their commutative counterparts. The fuzzy sphere is unstable and suffers collapse in a phase transition to Yang-Mills matrix models (topology change or geometric transition) whereas the noncommutative pseudo-sphere can sustain black hole configurations (by including a dilaton field) and also suffers collapse in the form of the information loss process (quantum gravity transition).

Explicitly, the noncommutative ${\rm AdS}^2_{\theta}$ is given by the following embedding and commutation relations

\begin{eqnarray}
  -\hat{X}_1^2+\hat{X}_2^2+\hat{X}_{3}^2=-R^2~,~[\hat{X}^a,\hat{X}^b]=i\kappa \epsilon^{ab}~_c\hat{X}^c.\label{well2}
\end{eqnarray}

\begin{eqnarray}
    \hat{X}^a=\kappa K^a.\label{well3}
\end{eqnarray}
The $K^a$ are the generators of the Lie group $SO(1,2)=SU(1,1)/\mathbb{Z}_2$ in the irreducible representations of the Lie algebra $[K^a,K^b]=i\epsilon^{ab}~_cK^c$  given by the discrete series $D_k^{\pm}$ with $k=\{1/2,1,2/3,2,3/2,...\}$ \cite{barg}.


By substituting the solution (\ref{well3}) in the constraint  equation in (\ref{well2}) the value of the deformation  parameter $\kappa$ is found to be  quantized in terms of the $su(1,1)$ pseudo-spin quantum number $j\equiv k-1$ as follows:
\begin{eqnarray}
\frac{R^2}{\kappa^2}=k(k-1).\label{rel1}
\end{eqnarray}
The commutative limit is then defined by $\kappa\longrightarrow 0$ or $k\longrightarrow \infty$.

The fact that we have for ${\rm AdS}^2$ two disconnected one-dimensional boundaries makes this case very different from higher dimensional anti-de Sitter spacetimes and is probably what makes the ${\rm AdS}^2$/${\rm CFT}_1$ correspondence the most mysterious case among all examples of the AdS/CFT correspondence. For example, see \cite{Strominger:1998yg}.

In \cite{Ydri:2021xpw} the difficulty of the  ${\rm AdS}^2$/${\rm CFT}_1$ correspondence is traced instead to the fact that the conformal quantum mechanics residing at the boundary is only quasi-conformal and as a consequence the theory in the bulk is only required to be quasi-AdS. In other words, we will take the opposite view and start or assume that the ${\rm CFT}_1$ theory on the boundary is really given by the dAFF conformal quantum mechanics \cite{deAlfaro:1976vlx,Chamon:2011xk}.

In the dAFF conformal quantum mechanics the $so(2,1)=su(1,1)$ Lie algebra generators $P$ (translations), $D$ (scale transformations) and $K$ (special conformal transformations) are realized in terms of a single degree of freedom $q(t)$ with scaling dimension $\Delta$. The corresponding conjugate momentum is given by $p(t)=\dot{q}(t)$ and we start from the standard Heisenberg algebra
\begin{eqnarray}
  [q,p]=i.
\end{eqnarray}
The $so(2,1)$ Lie algebra is given explicitly by 
\begin{eqnarray}
[D,P]=-iP~,~[D,K]=iK~,~[K,P]=-2iD.
\end{eqnarray}
A straightforward calculation gives then
\begin{eqnarray}
  P=\frac{p^2}{2}+V(q)~,~V(q)=\frac{g}{q^2}.
\end{eqnarray}
\begin{eqnarray}
    D=tP+\frac{\Delta}{2}(qp+pq)~,~\Delta=-\frac{1}{2}.
\end{eqnarray}
\begin{eqnarray}
    K=-t^2P+2tD+\frac{1}{2}q^2.
\end{eqnarray}
The coupling constant $g$ is completely fixed by conformal invariance. Indeed, we compute the Casimir operator

\begin{eqnarray}
  -C&=&\frac{1}{2}(PK+KP)-D^2=\frac{g}{2}-\frac{3}{16}\equiv \mp k(k-1).
\end{eqnarray}
The minus sign corresponds to Lorentzian  ${\rm AdS}^2$ (relevant to the dAFF quantum mechanics) whereas the plus sign corresponds to Euclidean  ${\rm AdS}^2$ (relevant to the noncommutative ${\rm AdS}^2_{\theta}$).

It is clearly observed that the Lorentz group $SO(1,2)$ is the fundamental unifying symmetry structure underlying the following commutative, noncommutative and quantum spaces:
\begin{enumerate}
\item The ${\rm AdS}^2$ spacetime (classical geometry and gravity).
\item The noncommutative ${\rm AdS}^2_{\theta}$ space (or first quantization of the geometry).
\item The quantum theory on the boundary (conformal quantum mechanics).
\item The IKKT-type Yang-Mills matrix models defining quantum gravity (or second quantization of the geometry).
  \end{enumerate}
In particular, the algebra of quasi-primary operators on the boundary is seen to be a subalgebra of the operator algebra of  noncommutative ${\rm AdS}^2_{\theta}$. This leads us to the conclusion/conjecture that the theory in the bulk must be given by noncommutative geometry and not by classical gravity, i.e. it is given by ${\rm AdS}^2_{\theta}$ and not by ${\rm AdS}^2$.  We end up therefore with an ${\rm AdS}^2_{\theta}$/${\rm CFT}_1$ correspondence where the ${\rm CFT}_1$ is given by the dAFF conformal quantum mechanics. See  \cite{Ydri:2021xpw}  for more detail.

This is then a correspondence or duality between quantum mechanics (${\rm QM}$) on the boundary and noncommutative geometry (${\rm NCG}$) in the bulk which provides a concrete model for the  ${\rm AdS}^{d+1}/{\rm CFT}_d$ correspondence \cite{Maldacena:1997re} in one dimension (see figure $1$).

In summary, the main argument underlying this ${\rm QM}/{\rm NCG}$ duality consists in the following main observations:
\begin{enumerate}
\item Both noncommutative ${\rm AdS}^2_{\theta}$ and the dAFF conformal quantum mechanics enjoy the same symmetry structure given by the group $SO(1,2)$. However, dAFF conformal quantum mechanics is only quasi-conformal in the sense that there is neither an invariant vacuum state nor strictly speaking primary operators  \cite{deAlfaro:1976vlx,Chamon:2011xk}. Analogously, noncommutative  ${\rm AdS}^2_{\theta}$ is only quasi-AdS as it approaches ${\rm AdS}^2$ only at large distances (commutative limit).
\item Asymptotically ${\rm AdS}^2_{\theta}$ is an ${\rm AdS}^2$ spacetime, i.e. it has the same boundary \cite{Pinzul:2017wch}. And furthermore the algebra of quasi-primary operators on the boundary (which defines in the same time the geometry of the boundary and the dAFF quantum mechanics) is in some sense a subalgebra of the operator algebra of noncommutative  ${\rm AdS}^2_{\theta}$  \cite{Ydri:2021xpw}.
\item Metrically the Laplacian operator on the noncommutative ${\rm AdS}^2_{\theta}$ shares the same spectrum as the Laplacian operator on the commutative ${\rm AdS}^2$ spacetime \cite{Jurman:2013ota}. The boundary correlation functions computed using the quasi-primary operators reproduces the bulk ${\rm AdS}^2$ correlation functions  \cite{Chamon:2011xk}.
\end{enumerate}

\begin{figure}[htbp]
\includegraphics[width=8.0cm,angle=0]{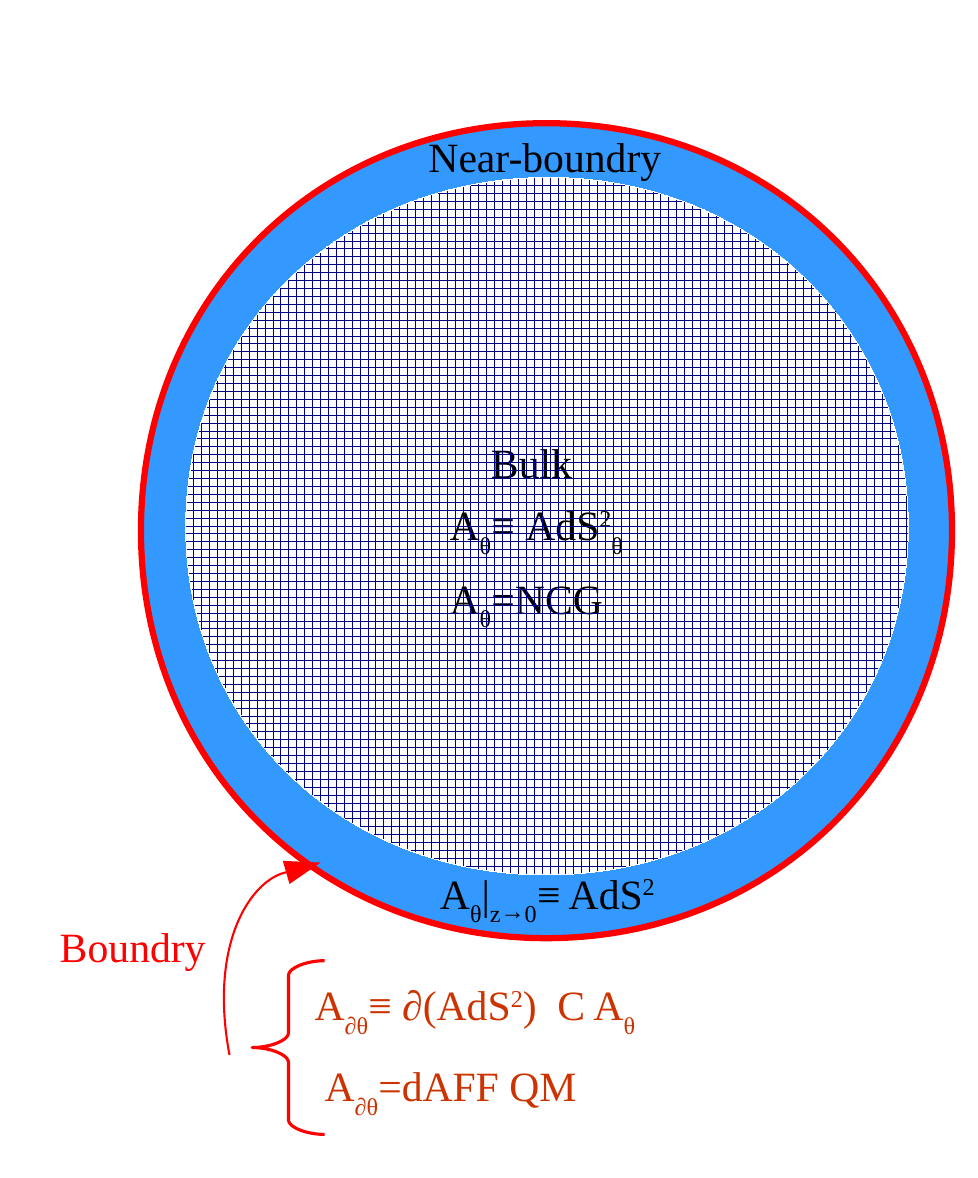}
\caption{The QM/NCG correspondence: The bulk is given by the algebra associated with noncommutative ${\rm AdS}^2_{\theta}$ while the boundary is given by a subalgebra thereof. The behavior near-boundary is precisely that of commutative ${\rm AdS}^2$. }\label{QMNCG}
\end{figure}

\section{Noncommutative black holes and matrix models}

The matrix model describing quantum gravitational fluctuations about the noncommutative ${\rm AdS}^2_{\theta}$ black hole is given by a $4-$dimensional Yang-Mills matrix model with mass deformations which are invariant only under the subgroup of $SO(1,2)$ pseudo-rotations around $D_4\equiv \Phi$ (dilaton field). This model is given explicitly by the action \cite{Ydri:2021qer}

\begin{eqnarray}
S[D]&=&N {\rm Tr}\bigg(-\frac{1}{4}[D_{\mu},D_{\nu}][D^{\mu},D^{\nu}]+2i\kappa D^2[D^1,D^3]+\beta D_aD^a\nonumber\\
&+&2i\kappa_1\hat{\Phi}[D^1,D^3]-\frac{1}{2}\kappa_1^2D_1D^1-\frac{1}{2}\kappa_1^2D_3D^3\bigg).\label{mp}
\end{eqnarray}
We have then two parameters $\kappa$ and $\kappa_1=2\pi T\kappa$ or equivalently $\kappa$ and $T$. We will fix the value of the parameter $\kappa_1$ so that an increase in the actual temperature $T$ of the black hole corresponds to a decrease in the noncommutativity parameter (or inverse gauge coupling constant) $\kappa$.   We are interested in the values $\kappa\neq 0$ and $\beta=0$.

The pure noncommutative ${\rm AdS}^2_{\theta}$ geometry appears as a classical (global) minimum of the matrix model (\ref{mp}) with $\kappa\neq 0$ and $\kappa_1=0$, i.e.  $T=0$.

Yang-Mills matrix models are generally characterized by two phases: 1) A Yang-Mills phase and 2) A geometric phase due to the Myers terms, i.e. the cubic terms in (\ref{mp}). 

The first cubic term in (\ref{mp}) is responsible for the condensation of the geometry (${\rm AdS}^2_{\theta}$ space) whereas the second cubic term in (\ref{mp}) is responsible for the stability of the noncommutative black hole solution in the ${\rm AdS}^2_{\theta}$ space.  There should then be two critical points ${\kappa}_{1 \rm cr}$ and  ${\kappa}_{2 \rm cr}$ corresponding to the vanishing of the tow cubic expectation values $\langle i {\rm Tr}  D^2[D^1,D^3]\rangle$ and $\langle i {\rm Tr} \Phi [D^1,D^3]\rangle$ respectively. By assuming now that $\kappa_1>\kappa$ we have ${\kappa}_{2 \rm cr}> {\kappa}_{1 \rm cr}$ and we have the following three phases (a gravitational phase, a geometric phase and a Yang-Mills phase):
\begin{itemize}
\item For $\kappa> \kappa_{2 \rm cr}$ we have a black hole phase where $\hat{\Phi}\propto \hat{X}_2$. This is the gravitational phase in which we have a stable noncommutative ${\rm AdS}^2_{\theta}$ black hole.
\item For ${\kappa}_{1 \rm}< \kappa < {\kappa}_{2 \rm}$ the second cubic term effectively vanishes and condensation is now only driven by the first cubic term. In other words, although we have in this phase $\hat{\Phi}\sim 0$ the three matrices $D_a$ still fluctuate about the noncommutative  ${\rm AdS}^2_{\theta}$ background given by the matrices  $\hat{X}_a$. This is the geometric phase.
\item As we keep decreasing the temperature, i.e. increasing $\kappa$ below $\kappa_1$ the   ${\rm AdS}^2_{\theta}$ background evaporates in a transition to a pure Yang-Mills matrix model. In this Yang-Mills phase, all four matrices become centered around zero. 
\end{itemize}
This is the Hawking process as it presents itself in the Yang-Mills matrix model, i.e. as an exotic line of discontinuous transitions with a jump in the entropy, characteristic of a 1st order transition, yet with divergent critical fluctuations and a divergent specific heat characteristic of a 2nd order transition.

An alternative way of recasting Hawking process as a phase transition starts from  dilaton gravity in two dimensions which can be rewritten as  a non-linear sigma model \cite{Cavaglia:1998xj}. For constant dilaton configurations, i.e. for $\Phi={\rm constant}$ this action reduces to the potential term \cite{Ydri:2021qer}
\begin{eqnarray}
V&=&m^2\int d^2x\sqrt{{\rm det} g}\bigg(M-N(\Phi)\bigg)^2~,~N(\Phi)=\int^{\Phi}d\Phi^{\prime} V(\Phi^{\prime}).
\end{eqnarray}
The limit $m^2\longrightarrow\infty$ is understood.  The mass functional $M$ is constant on the classical orbit given by the ADM mass of the  ${\rm AdS}^2$ black hole, viz $M=2\pi^2T^2/\Lambda$ \cite{Cadoni:2000wa}. For the case of the Jackiw-Teitelboim action we have $V=2\Lambda^2{\Phi}$ and $N=\Lambda^2{\Phi}^2$. The noncommutative analogue of this Jackiw-Teitelboim  potential term is a real quartic matrix model \cite{Shimamune:1981qf}
\begin{eqnarray}
V=N_{\rm H}{\rm Tr}(\mu\hat{\Phi}^2+g\hat{\Phi}^4)~,~N_{\rm H}\equiv 2k-1.
\end{eqnarray}
The regularization issues are discussed in the next section. The parameters $\mu$ and $g$ are given explicitly by
\begin{eqnarray}
\mu=-2\pi M.(\frac{m^2}{k^2})~,~g=\pi\Lambda^2.(\frac{m^2}{k^2}).
\end{eqnarray}
The one-cut solution (disordered phase) in which the dilaton field $\hat{\Phi}$ fluctuates around zero occurs for all values $M\leq M_*$ where

\begin{eqnarray}
M_*=\frac{\Lambda}{\sqrt{\pi}}\frac{k}{m}.
\end{eqnarray}
The large mass limit $m^2\longrightarrow \infty$ is then required to be correlated with the commutative limit $k\longrightarrow \infty$. This critical value $M_*$ is consistent with the black hole mass $M=2\pi^2T^2/\Lambda$. In other words, the dilaton  field is zero when the mass parameter $M$ of the matrix model is lower than the black hole mass, i.e. the black hole has already evaporated and we have only pure ${\rm AdS}^2_{\theta}$ space in this phase.

Above the line $M=M_*$ we have a two-cut solution (non-uniform ordered phase) where the dilaton field $\hat{\Phi}$ is proportional to the idempotent matrix $\gamma$ (which satisfies $\gamma^2={\bf 1}$).  In this two-cut phase we have therefore a non-trivial dilaton field corresponding to a stable ${\rm AdS}^2_{\theta}$  black hole. The transition at $M=M_*$ between the the one-cut (${\rm AdS}^2_{\theta}$ space) and two-cut phases (${\rm AdS}^2_{\theta}$ black hole) is third order.

\section{Phase Structure of the Noncommutative ${\rm AdS}^2_{\theta}\times \mathbb{S}^2_N$}

The quantum gravitational fluctuations about the noncommutative geometry of the fuzzy sphere $\mathbb{S}^2$, the noncommutative pseudo-sphere $\mathbb{H}^2$ and the noncommutative near-horizon geometry ${\rm AdS}^2_{\theta}\times\mathbb{S}^2_N$ are given by IKKT-type Yang-Mills matrix models with additional cubic Myers terms which are given explicitly by the actions 

\begin{eqnarray}
S_{\rm S}[C]&=&N_{\rm S} {\rm Tr}_{\rm S}L_{\rm S}[C]~,~\mathbb{S}^2.\label{action1}
\end{eqnarray}
\begin{eqnarray}
S_{\rm H}[D]&=&N_{\rm H} {\rm Tr}_{\rm H}L_{\rm H}[D]~,~\mathbb{H}^2.\label{action2}
\end{eqnarray}
\begin{eqnarray}
S_{\rm HS}[D,C]&=&N_{\rm H} N_{\rm S}{\rm Tr}_{\rm H}{\rm Tr}_{\rm S}\bigg(L_{\rm S}[C]+L_{\rm H}[D]-\frac{1}{4}[D_a,C_b][D^a,C_b]\bigg)~,~\mathbb{H}^2\times\mathbb{S}^2.\label{action3}\nonumber\\
\end{eqnarray}
The Lagrangian terms are given in terms of three $(2l+1)\times (2l+1)$ Hermitian matrices $C_a$  (where $N_{\rm S}=2l+1\equiv N$) and three Hermitian operators $D_a$ by the equations
\begin{eqnarray}
L_{\rm S}[C]&=&-\frac{1}{4}[C_a,C_b]^2+\frac{2i}{3}\alpha \epsilon_{abc}C_aC_bC_c+\beta_S C_a^2.
\end{eqnarray}
\begin{eqnarray}
L_{\rm H}[D]&=&-\frac{1}{4}[D_a,D_b][D^a,D^b]+\frac{2i}{3}\kappa f_{abc}D^aD^bD^c+\beta_H D_aD^a.
\end{eqnarray}
The coefficients $\epsilon_{abc}$ and  $f_{abc}$ provide the structure constants of the rotation group $SO(3)=SU(2)/\mathbb{Z}_2$ and the pseudo-rotation group  $SO(1,2)=SU(1,1)/\mathbb{Z}_2$  respectively. 

The solutions of the equations of motion which follow from the actions (\ref{action1}), (\ref{action2}) and (\ref{action3}) are precisely given by the fuzzy  sphere $\mathbb{S}^2_N$, the noncommutative pseudo-sphere $\mathbb{H}^2_{\theta}$ and the noncommutative near-horizon geometry ${\rm AdS}^2_{\theta}\times\mathbb{S}^2_N$ configurations. These noncommutative configurations are given explicitly by
\begin{eqnarray}
C_a=\phi_S L_a~,~\phi_S=\alpha\varphi_S.\label{S}
\end{eqnarray}
\begin{eqnarray}
D_a=\phi_H K_a~,~\phi_H=\kappa\varphi_H.\label{H}
\end{eqnarray}
\begin{eqnarray}
C_a=\phi_S L_a\otimes {\bf 1}_H~,~D_a={\bf 1}_S\times\phi_H K_a.\label{HS}
\end{eqnarray}
These matrix models  (\ref{action1}), (\ref{action2}) and (\ref{action3}) exhibit emergent geometry transitions from a geometric phase (the sphere, the pseudo-sphere and the near-horizon geometry of a Reissner-Nordstrom black hole) to a pure Yang-Mills matrix phase with no background geometrical structure.

This fundamental result is confirmed in the case of the sphere by Monte Carlo simulations of the Euclidean Yang-Mills matrix model (\ref{action1}). But in the other cases we have only at our disposal the effective potential. Indeed, the Monte Carlo method can not be applied to the matrix models (\ref{action2}) and (\ref{action3}) in their current form for two main reasons. First, the Lorentzian signature of the embedding spacetime  (these noncommutative spaces are in fact branes residing in a larger spacetime). Second, the Hilbert spaces ${\cal H}_k$ corresponding to the noncommutative operator algebras are actually infinite-dimensional (which is required by the underlying $SO(1,2)$ symmetry).

However, the matrix models (\ref{action2}) and (\ref{action3}) can be regularized by means of a simple cutoff which consists in keeping only the $2k-1$ states in the Hilbert spaces turning therefore the infinite-dimensional operator algebras into finite-dimensional matrix algebras. By assuming also that the ${\rm AdS}$ spacetime has the same radius as the sphere we have the natural identification 

\begin{eqnarray}
N_{\rm H} =2k-1\equiv N.\label{Hreg}
\end{eqnarray}
The one-loop effective potential around the  pseudo-sphere background (\ref{H}) is computed from the action (\ref{action2}) using the background field method. We obtain the result (with $\tilde{\kappa}^4=4\kappa^4 k(k-1)$ and $\tau_H=-{\beta_H}/{\kappa^2}$)
\begin{eqnarray}
\frac{2V_H}{N^2}
&=&\tilde{\kappa}^4 \bigg[\frac{1}{4}\varphi_H^4-\frac{1}{3}\varphi_H^3+\frac{1}{2}\tau_H\varphi_H^2\bigg]+\log\varphi_H^2~,~T_H=\frac{1}{\tilde{\kappa}^4}=g_H^2.\label{effe2}
\end{eqnarray}
The one-loop effective potential around the sphere background (\ref{S}) is obtained from this result with the substitutions $k\longrightarrow-l$, $\varphi_H\longrightarrow\varphi_S$, $\kappa\longrightarrow\alpha$, $\beta_H\longrightarrow -\beta_S$, $\tau_H\longrightarrow\tau_S$, $g_H\longrightarrow g_S$ and $T_H\longrightarrow T_S$.


The description of the phase structure of the noncommutative pseudo-sphere $\mathbb{H}^2_{\theta}$ using the effective potential (\ref{effe2}) is therefore formally identical to the fuzzy sphere case treated in detail in \cite{CastroVillarreal:2004vh,Azuma:2004zq,Delgadillo-Blando:2007mqd, Delgadillo-Blando:2012skh}.

The sphere/pseudo-sphere solution $\varphi_{S,H}=\varphi_-(\tau_{S,H})\neq 0$ is found to be stable only in the regime $0<\tau_{S,H}<2/9$. This is the fuzzy or noncommutative geometry phase. The Yang-Mills solution  $\varphi_{S,H}=\varphi_0=0$ (matrix phase) is found to be the ground state or global minimum of the system in the regime $2/9<\tau_{S,H}<1/4$. In fact,  the coexistence curve between the geometric phase (low temperature $T_H$) and the matrix phase (high temperature $T_H$) asymptotes to the line $\tau_{S,H}=2/9$.

We turn now to the case of the  noncommutative near-horizon geometry  ${\rm AdS}^2_{\theta}\times\mathbb{S}^2_N$. We can immediately state the possible phases of  the noncommutative near-horizon geometry  ${\rm AdS}^2_{\theta}\times\mathbb{S}^2_N$ in the space $(\tau_S,\tau_H,\bar{\alpha},\bar{\kappa})$ as follows \cite{Ydri:2021vxs}
\begin{itemize}
\item A Yang-Mills matrix phase with no background geometrical structure which is expected at high temperature (both $\bar{\alpha}$ and $\bar{\kappa}$ approach zero) or $2/9<\tau_{S,H}<1/4$.
\item A $2-$dimensional geometric fuzzy sphere phase ($\bar{\alpha}$  approaches infinity, $\bar{\kappa}$ approaches zero and  $0<\tau_{S,H}<2/9$).
\item A $2-$dimensional geometric noncommutative pseudo-sphere phase ($\bar{\alpha}$  approaches zero, $\bar{\kappa}$ approaches infinity and  $0<\tau_{S,H}<2/9$).
\item A $4-$dimensional geometric noncommutative near-horizon geometry  ${\rm AdS}^2_{\theta}\times\mathbb{S}^2_N$ phase at low temperature (both $\bar{\alpha}$ and $\bar{\kappa}$ approach infinity and  $0<\tau_{S,H}<2/9$).
\end{itemize}
In order to simplify the calculation of the one-loop effective potential around the near-horizon geometry background (\ref{HS}) we will consider the following case:

  \begin{itemize}
    \item First, in order to have a single unified temperature $T_{\rm HS}$ we must have a single unified gauge coupling constant $g_{\rm HS}^2=1/T_{\rm HS}$, viz
\begin{eqnarray}
  \alpha=\kappa.\label{beta0}
\end{eqnarray}
  \item The geometric quantum entanglement between the sphere and the pseudo-sphere sectors (through the third term in the action (\ref{action3})) is essential for the emergence of a four-dimensional space.  However, there exists a special case where this geometric quantum entanglement can be  partially removed while keeping the background emergent geometry four-dimensional. This is given by the case
    \begin{eqnarray}
\tau_S=\tau_H\iff \beta_S=-\beta_H.\label{beta}
    \end{eqnarray}
    \item With this choice (\ref{beta}) one can check that the order parameters $\varphi_S$ and $\varphi_H$ must solve identical equations of motion. A simplified model  is then obtained by setting 
\begin{eqnarray}
\varphi_S=\varphi_H\equiv\varphi.\label{beta1}
\end{eqnarray}
    \end{itemize}
The one-loop effective potential around the near-horizon geometry background (\ref{HS})  computed from the action (\ref{action3}) using the background field method is then given by

\begin{eqnarray}
\frac{V_{HS}}{2N_{\rm S}^2N_{\rm H}^2}
&=&2\bar{\kappa^4}\bigg[\frac{1}{4}\varphi^4-\frac{1}{3}\varphi^3+\frac{1}{2}\tau_H\varphi^2\bigg]+\log\varphi^2~,~\bar{\kappa}^4=\frac{\tilde{\kappa}^4}{4N}.\label{quantum}\nonumber\\
\end{eqnarray}
This effective potential is of the same form as the pseudo-sphere effective potential (\ref{effe2}) with the substitution $\tilde{\kappa}^4\longrightarrow 2\bar{\kappa^4}$.

In the special case given by  (\ref{beta0}), (\ref{beta}) and (\ref{beta1}) the $2-$dimensional geometric phases disappear and we end up with a single phase transition from a noncommutative near-horizon geometry  ${\rm AdS}^2_{\theta}\times\mathbb{S}^2_N$ phase to a Yang-Mills matrix phase as the temperature $T_{\rm HS}=1/\bar{\kappa}^4=g_{\rm HS}^2$ is increased.

\end{document}